\documentclass[10pt,conference]{IEEEtran}
\IEEEoverridecommandlockouts
\usepackage{cite}
\usepackage{amsmath,amssymb,amsfonts}
\usepackage[ruled,vlined,linesnumbered]{algorithm2e}
\usepackage{amsthm}
\newtheorem{claim}{Claim}
\usepackage{subcaption}\usepackage{graphicx}
\usepackage{textcomp}
\usepackage[]{amsmath, amssymb}
\usepackage[]{quantikz}
\usepackage{float}
\usepackage{xcolor}
\def\BibTeX{{\rm B\kern-.05em{\sc i\kern-.025em b}\kern-.08em
    T\kern-.1667em\lower.7ex\hbox{E}\kern-.125emX}}
\begin{document}

\title{A Fully Quantum Algorithm for Image Edge Detection\\
}

\author{\IEEEauthorblockN{Fred Sun}
\IEEEauthorblockA{\textit{Institute for Quantum Computing, University of Waterloo}, Waterloo, ON, Canada \\
f4sun@uwaterloo.ca}
}

\maketitle

\begin{abstract}
This work introduces a novel quantum algorithm for gradient-based edge detection that operates entirely within the quantum circuit model. Grayscale images are encoded using the Novel Enhanced Quantum Representation (NEQR), allowing exact arithmetic on pixel intensities. Directional gradients are computed by generating superpositions of neighboring pixels via cyclic shift operations and performing subtraction with an exact quantum arithmetic circuit. To refine accuracy, we introduce a direction-aware shifting mechanism that aligns edges with the darker side of intensity transitions. Our novel Quantum Partitioning Algorithm enables efficient in-place thresholding of edge candidates. This work exhibits polynomial-time improvements and optimizes the ancilla count compared to previous NEQR-based quantum edge detection algorithms. These results demonstrate a resource-efficient and fully quantum approach to edge detection, highlighting a practical quantum advantage in image processing.
\end{abstract}


\section{\label{sec:level1}Introduction}

Edge detection is a fundamental process in image processing, with many applications in computer vision, image compression, and feature extraction. These algorithms identify discontinuities in pixel intensities, which often correspond to object boundaries. Classical edge detection methods, such as the Sobel and Canny filters, achieve high accuracy by estimating intensity gradients via convolution with local kernels \cite{Sobel, Canny}. However, despite their effectiveness, these methods are computationally expensive, particularly for high-resolution images or large datasets. Their inherent reliance on sequentially processing each pixel results in slow and increasingly impractical runtime as input size scales. For time-critical applications, such as autonomous driving and real-time medical imaging, these computational delays are especially problematic, motivating the need for faster and more scalable alternatives. 

Quantum computing leverages the quantum principles of superposition and entanglement to process information in a fundamentally different way than classical systems. Superposition allows a quantum bit (qubit) to exist in a linear combination of both the \ket{0} and \ket{1} states simultaneously until measured. Consequently, an $n$-qubit system can encode $2^n$ states in parallel, allowing a quantum computer to operate on an exponentially large state space with just one computational step. For image processing, this intrinsic parallelism allows operations to be performed simultaneously on all encoded pixel positions and intensities. In addition, entanglement enables qubits to share non-classical correlations, allowing for more complex joint operations across the image data. These quantum properties offer significant advantages in computational complexity, making quantum computing a promising tool for scalable image analysis \cite{yansurvey}\cite{nielsenchuang}.

Prior work in quantum image processing (QIMP) has introduced several image representation models, such as the Flexible Representation of Quantum Images (FRQI) and Novel Enhanced Quantum Representation (NEQR) \cite{yaoqhed, Zhang_Lu_Gao_Wang_2013}. In this work, we use NEQR for its ability to encode each pixel's intensity value deterministically in the computational basis, enabling localized quantum operations such as quantum arithmetic.

In this article, we present a novel quantum algorithm for gradient-based edge detection that operates entirely within the quantum circuit model on NEQR-encoded grayscale images. The algorithm comprises three main components: gradient computation, direction-aware edge shifting, and efficient quantum thresholding. Directional gradients are calculated by generating superpositions of neighboring pixels using cyclic shift operations and performing subtraction with an exact quantum arithmetic circuit. To improve localization accuracy, the direction-aware shifting mechanism aligns edges with the darker side of intensity transitions. Edge classification is achieved using a Quantum Partitioning Algorithm, which deterministically separates edges from non-edges in a single call to a Grover-style phase oracle implemented in logarithmic circuit depth. By performing all stages fully in the quantum domain, the algorithm eliminates classical post-processing overhead while leveraging quantum parallelism for scalable performance. In practical scenarios, the algorithm achieves exponential compression of the output edge map, exhibits $O(n+q)$ time complexity for most practical cases in image processing, and requires only $3q+O(1)$ computational qubits for a $2^n \times 2^n$ image with $2^q$ grayscale levels, demonstrating a resource-efficient and fully quantum approach to edge detection.

The article is organized as follows. Section~\ref{sec:background} reviews related work in quantum image encoding, quantum arithmetic, and previous quantum edge detection methods. Section~\ref{sec:method} details our proposed quantum edge detection algorithm, including a high-level overview and implementation, and a detailed description of each submodule. Section~\ref{sec:results} analyzes the algorithm’s time and space complexity, compares it with prior works, and presents simulation-based experimental results demonstrating its effectiveness. Finally, Section~\ref{sec:conclusion} summarizes the findings and discusses potential extensions and applications of the proposed approach.

\section{Background and Related Work} \label{sec:background}

Various quantum image processing techniques for image representation and analysis tasks, such as edge detection, have been proposed. This section reviews relevant prior works in quantum image encoding, Quantum Hadamard Edge Detection (QHED), and NEQR-based edge detection methods that apply Sobel-like filters using cyclic pixel shifts. 

\subsection{Quantum Image Encoding}

Several quantum image representation models have been studied, each offering various trade-offs in terms of storage size, encoding time complexity, and suitability for various tasks. The Quantum Probability Image Encoding (QPIE) method represents grayscale image data using amplitude encoding \cite{yaoqhed}. Given a grayscale image of size $2^n \times 2^n$ with normalized pixel intensities $I(x,y)$, QPIE represents the image as:
\begin{align}
\ket{\psi_{\text{QPIE}}} = \sum_{x=0}^{2^{n}-1}\sum_{y=0}^{2^{n}-1}I(x,y)\ket{x}\ket{y}
\end{align} 
This encoding scheme is the most qubit-efficient, only requiring $2n$ qubits to encode an $n \times n$ image. However, because the intensity values are encoded into amplitudes, accessing this information requires measurement, which is inherently probabilistic, and collapses the quantum state. This limits QPIE's use for applications that require deterministic manipulation or access to the pixel intensity values.

By contrast, the Novel Enhanced Quantum Representation (NEQR) encodes grayscale values directly in the computational basis rather than as an amplitude \cite{Zhang_Lu_Gao_Wang_2013}. An image of size $2^n \times 2^n$ with $2^q$ intensity values is represented as: 
\begin{align}\ket{\psi_{\text{NEQR}}} = \sum_{x=0}^{2^{n}-1}\sum_{y=0}^{2^{n}-1}\ket{x}\ket{y}\ket{I(x,y)}
\end{align}
NEQR requires more qubits, $2n+q$, but provides deterministic and direct access to both the position and intensity values of each pixel, allowing for localized quantum operations, such as quantum arithmetic. Amankwah et al. introduced an efficient method to encode a $2^n \times 2^n$ image with $q$ color values in NEQR with a time complexity of $O(q\times2^n)$ \cite{Amankwah_2022}. This ability to encode and manipulate intensity values at the state level is critical to the design of this work’s edge detection algorithm, which relies on pixel-wise gradient calculations, thresholding, and conditional shifting, making NEQR the ideal choice for this work.

\subsection{Quantum Arithmetic}

Quantum arithmetic circuits provide the foundation for this work, enabling essential operations such as gradient computation and threshold-based decision making entirely within a quantum circuit, without requiring intermediate measurements. The ability to execute these operations while maintaining quantum coherence is essential for maintaining quantum parallelism and avoiding the computational overhead associated with switching between quantum and classical data.

\subsubsection{Addition}
A core method of quantum arithmetic is quantum addition, which can be extended to other arithmetic operations such as subtraction. Given the input states $\ket{x}\ket{y}$, the quantum adder outputs $\ket{x}\ket{x+y}$. We will be using the Quantum Ripple Carry Adder, proposed by Cuccaro et al. \cite{cuccaro2004newquantumripplecarryaddition}.

\begin{figure}[h]
\centering
\resizebox{\columnwidth}{!}{
\begin{quantikz}[column sep=0.5cm] 
\lstick{$\ket{c_0}$} & \gate[wires=3][0.5cm]{\text{MAJ}} & \qw & \qw & \qw& \qw& \qw &\gate[wires=3][0.5cm]{\text{UMA}} &\push{\text{ }\ket{0}}\\
\lstick{$\ket{b_0}$} &                         & \qw & \qw & \qw& \qw& \qw &\qw &\push{\text{ }\ket{s_0}}\\
\lstick{$\ket{a_0}$} &                         & \gate[wires=3][0.5cm]{\text{MAJ}}                    & \qw & \qw&  \qw &\gate[wires=3][0.5cm]{\text{UMA}}& \qw&\push{\text{ }\ket{a_0}}\\
\lstick{$\ket{b_1}$} & \qw                     &                     & \qw & \qw& \qw& \qw& \qw& \push{\text{ }\ket{s_1}}\\
\lstick{$\ket{a_1}$} & \qw     & \qw                 & \gate[wires=3][0.5cm]{\text{MAJ}} & \qw& \gate[wires=3][0.5cm]{\text{UMA}}& \qw &\qw &\push{\text{ }\ket{a_1}}\\
\lstick{$\ket{b_2}$} &                         & \qw                 & \qw & \qw& \qw& \qw &\qw& \push{\text{ }\ket{s_2}}\\
\lstick{$\ket{a_2}$} & \qw                     & \qw                 & \qw & \ctrl{1}& \qw& \qw &\qw& \push{\text{ }\ket{a_2}}\\
\lstick{$\ket{z}$} & \qw & \qw & \qw & \targ{} & \qw& \qw & \qw&\push{\text{ }\ket{z \oplus s_3}}
\end{quantikz}
}
\caption{3-qubit QRCA circuit}
\label{fig:qrca}
\end{figure}
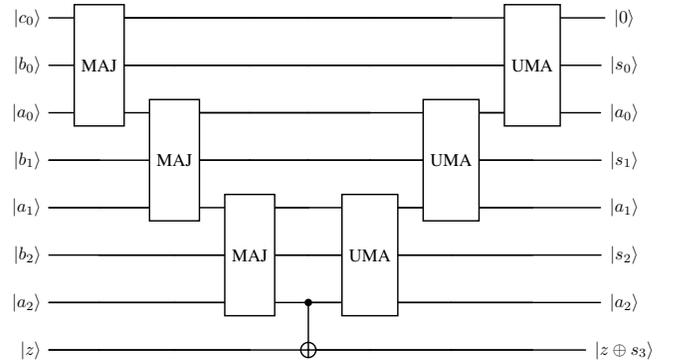

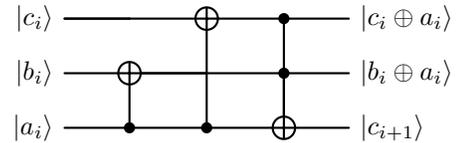
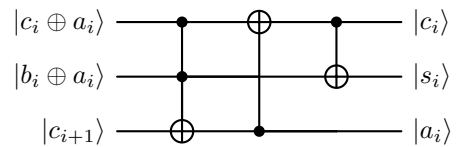
\begin{figure}[h]
\centering
\begin{subfigure}{0.4\textwidth}
\centering
\begin{quantikz}[column sep=0.7cm, row sep=0.4cm]
\lstick{$\ket{c_i}$} & \qw & \targ{} & \ctrl{2} & \rstick{$\ket{c_i \oplus a_i}$} \\
\lstick{$\ket{b_i}$} & \targ{} & \qw & \ctrl{1} & \rstick{$\ket{b_i \oplus a_i}$}\\
\lstick{$\ket{a_i}$} & \ctrl{-1} & \ctrl{-2} & \targ{} & \rstick{$\ket{c_{i+1}}$} \\
\end{quantikz}
\caption{Majority (MAJ)}
\label{fig:maj}
\end{subfigure}

\par\medskip

\begin{subfigure}{0.4\textwidth}
\centering
\begin{quantikz}[column sep=0.7cm, row sep=0.4cm]
\lstick{$\ket{c_i \oplus a_i}$} & \ctrl{2} & \targ{} & \ctrl{1} & \rstick{$\ket{c_i}$} \\
\lstick{$\ket{b_i \oplus a_i}$} & \ctrl{1} & \qw & \targ{} & \rstick{$\ket{s_i}$}\\
\lstick{$\ket{c_{i+1}}$} & \targ{} & \ctrl{-2} & \qw & \rstick{$\ket{a_i}$} \\
\end{quantikz}
\caption{Unmajority-and-Add (UMA)}
\label{fig:uma}
\end{subfigure}

\caption{Submodules of the QRCA}
\label{fig:majuma}
\end{figure}

The QRCA is analogous to the classical full-adder \cite{mano2017digital}, using reversible majority-3 (MAJ) gates to compute the carry-out, $c_{i+1} = \text{MAJ}(a_i,b_i,c_i) = (a_i b_i) \oplus (a_i c_i) \oplus (b_i c_i)$, and unmajority-and-sum (UMA) to produce the sum bits while reverting all intermediate calculations. Fig.~\ref{fig:qrca} and Fig.~\ref{fig:majuma} show the high-level overview of the QRCA circuit implementation.

For $n$-bit addition, the circuit depth is $O(n)$ with only one ancillary qubit required for the carry-out. 

\subsubsection{Subtraction}\label{subsec:subtraction}

A negative number is represented using the \textit{two's complement encoding}. To transform $y$ into $-y$, we flip all the bits of $y$, then add $1$. To implement this, we first apply an $X$ gate to all qubits of $\ket{y}$ to obtain $\overline{y} = \overline{y_{n-1}}\dots\overline{y_0}$. Then, addition by one is performed using a series of CNOT gates starting from the least significant qubit, targeting the next more significant qubit. This operation will be referred to as \texttt{S2C}.

We now implement the subtraction $x-y$ by reducing it to the addition of $x$ and the two's complement of $y$. Given registers $\ket{x}$ and $\ket{y}$, we compute
\[
\ket{x}\ket{y}
\;\longmapsto\;
\ket{x}\ket{x-y}
\]

The first step is to apply \texttt{S2C} to $\ket{y}$, transforming it to $\ket{-y}$. We then perform the QRCA with inputs $\ket{x}$ and $\ket{-y}$, which produces an output register $\ket{x-y}$.

It is clear that the asymptotic resource costs of this operation are the same as the QRCA.

\subsection{Quantum Hadamard Edge Detection}

Quantum Hadamard Edge Detection (QHED), introduced in 2016 by Yao et al., operates on grayscale images encoded in QPIE \cite{yaoqhed}. This method uses gradient approximation to predict the presence of edges using superposition and amplitude interference. Given a $2^{n/2} \times 2^{n/2}$ image, the vector form of a QPIE representation is:
\begin{align}
\begin{bmatrix} c_0 \\ 
c_1 \\ 
\vdots \\ 
c_{2^n-2} \\ 
c_{2^n-1} 
\end{bmatrix}
\end{align}

QHED then applies a Hadamard gate to the least significant qubit while doing nothing to the others, yielding:
\begin{align}
\frac{1}{\sqrt{2}}
\begin{bmatrix}
c_0 + c_1 \\
c_0 - c_1 \\
c_2 + c_3 \\
c_2 - c_3 \\
\vdots \\
c_{2^n-2} + c_{2^n-1} \\
c_{2^n-2} - c_{2^n-1}
\end{bmatrix}
\end{align}

In this result, it is observed that odd-indexed rows represent the intensity gradient of neighboring pixels, where a non-zero gradient indicates an edge. If the least significant qubit is measured in the $\ket{1}$ state, then the system collapses to the subspace containing the odd-indexed rows: 
\begin{align}
\begin{bmatrix}
c_0 - c_1 \\
c_2 - c_3 \\
\vdots \\
c_{2^n-2} - c_{2^n-1}
\end{bmatrix}
\end{align}
In this state, all amplitudes that evaluate to $0$ (i.e., where the neighboring pixel intensities are the same) are effectively filtered out of the system. The remaining states are those that correspond to the boundaries between pixels 0/1, 2/3, etc. To obtain the boundaries between the remaining pairs, an $n$-qubit amplitude permutation operation is performed on the input image state. The runtime of this algorithm is bounded by the amplitude permutation step, which can be implemented in $O(n^2)$ depth \cite{Shubha2024pex} \cite{Amankwah_2022}.

However, since the only information preserved after measurement is whether there was a non-zero intensity difference between neighboring pixels, QHED requires repeated sampling and relies on classical post-processing to reconstruct the edge map, limiting its efficiency and scalability. Furthermore, the direction of the intensity transitions (i.e., light to dark versus dark to light) is encoded in the sign of the amplitude, which is lost due to the nature of quantum measurements. 

An improved QHED proposed in 2024 by Shubha et al. applies classical post-processing to enhance accuracy and suppress noise in the output \cite{Shubha2024pex}. After measurement, thresholding, and edge alignment are performed on the classical data, which improves edge quality but introduces significant computational overhead and remains constrained by the probabilistic nature of amplitude-encoded images. These limitations motivate the development of a fully quantum approach that leverages NEQR encoding for deterministic intensity access and enables completely in-circuit edge detection without reliance on classical post-processing.

\subsection{Liu and Wang's NEQR Cyclic Shift Method}
Some research has already explored the use of NEQR encoding for quantum edge detection. Most of these methods are classically-inspired; using quantum circuits to mimic classical convolution filters such as the Sobel operator \cite{Sobel}. By using NEQR's ability to access pixel intensities, these approaches are able to perform pixel-wise arithmetic using shifted versions of the image. 

Liu and Wang’s method builds on this by introducing a cyclic shift operator to emulate the Sobel filter in the quantum domain \cite{Liu_2022}. Their approach prepares a quantum state containing ten pixels from the target pixel’s neighborhood in superposition. Eight of these pixels correspond to the immediate neighbors in each gradient direction, while the remaining two represent central reference values. The algorithm then applies quantum arithmetic to compute directional gradients for each of the eight neighboring windows. A maximum-selection circuit identifies the largest directional gradient magnitude, which is assigned as the pixel’s gradient strength.

A distinguishing feature of Liu and Wang’s work is the integration of post-processing within the quantum pipeline. They provide quantum implementations for non-maximum suppression, double-threshold detection, and edge tracking, effectively translating a complete classical edge-detection process into quantum circuits. This eliminates the need for intermediate measurements between gradient computation and edge refinement, preserving quantum coherence.

While the method demonstrates a comprehensive quantum mapping of a classical edge-detection pipeline, it requires maintaining multiple shifted image registers, one for each gradient direction, along with numerous comparator circuits. This design significantly increases both the qubit count and gate complexity. Furthermore, relying on storing each neighbor in a separate register presents scalability challenges for processing large images on Noisy Intermediate-Scale Quantum (NISQ) devices.

\section{Proposed Method} \label{sec:method}

\subsection{Algorithm Overview}

The proposed algorithm integrates cyclic-shifting gradient computation, direction-aware edge alignment, and thresholding into a fully quantum pipeline to output an edge-filtered image. Our method operates entirely in the quantum domain, avoiding intermediate measurements and classical processing, thus preserving quantum parallelism. Figure presents the high-level circuit diagram for the algorithm, and Table~\ref{tab:edge_steps} provides more in-depth state details for each step. 

\begin{figure*}[ht]
    \centering
    \includegraphics[width=\textwidth]{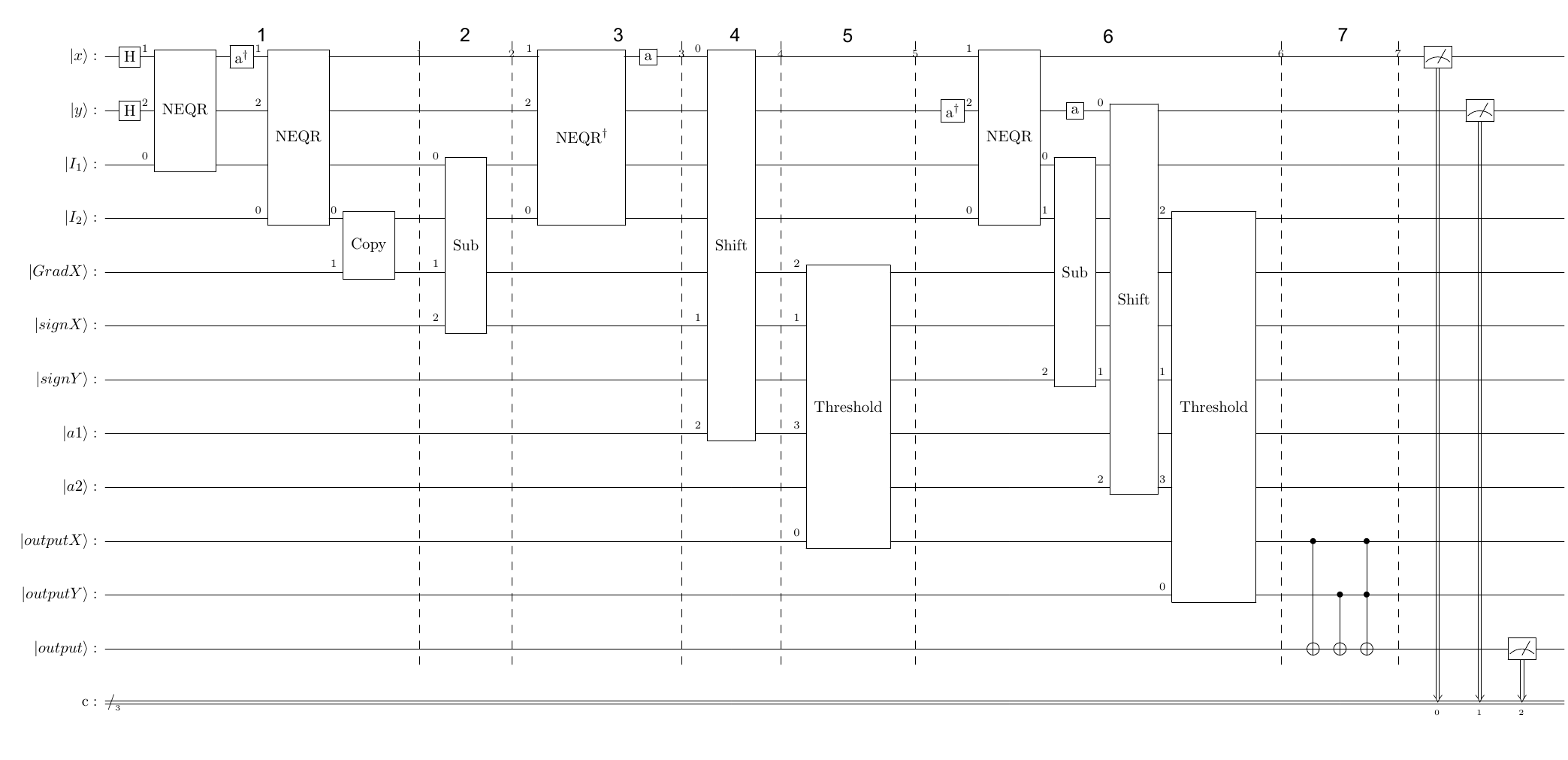}
    \caption{High-level circuit diagram of the proposed quantum edge detection algorithm. All input registers are initialized with $\ket{0}$}
    \label{fig:algo_circuit}
\end{figure*}

\textbf{Step 1: Encoding.}
The NEQR oracle produces an intensity value based on the pixel position registers. We create a superposition over all position states and apply the NEQR oracle to put all pixel information in parallel. The $x$-position register is then shifted up by 1, using the $a^{\dagger}$ operator, and the NEQR oracle is reapplied, but with a different target register. This creates a superposition over all horizontally neighboring pixel pairs with both intensities available in the same state.

\textbf{Step 2: Compute $x$ gradient.}
The intensity registers are subtracted to compute the $x$-gradient value for each pixel. 

\textbf{Step 3: Reset for $y$ gradient.}
Reset the $I_2$ register by applying the inverse NEQR oracle and shift the $x$ register back down using the $a$ operator, to be reused for y-direction gradient calculation.

\textbf{Step 4: Direction-Aware Edge Shifting.}
Edge positions are conditionally shifted toward regions of lower colour intensity, consistent with physical image boundaries.

\textbf{Step 5: Thresholding}
All pixels are separated via entanglement with an ancillary qubit, $\text{output}_x$, based on whether their intensity value exceeds an input threshold, $T$, or not.

\textbf{Step 6: Edge Detection in $y$ direction}
We repeat the same process for edge detection along the $y$ direction. The detected edges in the $y$ direction are marked by the $\text{output}_y$ qubit.

\textbf{Step 7: Output}
Apply $\text{OR}(\text{output}_x, \text{output}_y)$ to mark all pixels that will be present in the final edge map with the $\ket{1}$ state in the output qubit.

\begin{table*}
\caption{Quantum edge detection algorithm steps.}
\label{tab:edge_steps}
\begin{tabular}{c c}
\hline
\textbf{Step} & \textbf{Quantum State} \\
\hline
\\
1 &
$\ket{\psi_1} = \sum_{x,y} \ket{x+1}\ket{y} \ket{I(x,y)}\ket{I(x+1,y)}$ \\ \\
\hline
\\
2 &
$\ket{\psi_2} = \sum_{x,y} \ket{x+1}\ket{y} \ket{I(x,y)}\ket{I(x+1,y)} \ket{|\Delta I_x|}\ket{\text{sign}_{\Delta I_x}}$ \\
\\
\hline
\\
3 &
$\ket{\psi_3} = \sum_{x,y} \ket{x}\ket{y} \ket{I(x,y)}\ket{0}
\ket{|\Delta I_x|}\ket{\text{sign}_{\Delta I_x}}$ \\
\\
\hline
\\
4 &
$\begin{aligned}
\ket{\psi_4} &= \sum_{\text{non-shifted}} \Big(
\ket{xy} \ket{I(x,y)} \ket{0}
\ket{|\Delta I_x|} \ket{\text{sign}_{\Delta I_x} = 0} \ket{a_1=0} \Big)+ \sum_{\text{shifted}} \Big(
\ket{x}\ket{y} \ket{I(x,y)} \ket{0}
\ket{|\Delta I_x|} \ket{\text{sign}_{\Delta I_x} = 0} \ket{a_1=1} \\
&\quad + \ket{x+1}\ket{y} \ket{I(x+1,y)} \ket{0}
\ket{|\Delta I_x|} \ket{\text{sign}_{\Delta I_x} = 1} \ket{a_1=1} \Big)
\end{aligned}$ \\
\\
\hline
\\
5 &
$\begin{aligned}
\ket{\psi_5} &= \sum_{\text{non-shifted}} \Big(
\ket{xy} \ket{I(x,y)} \ket{0}
\ket{|\Delta I_x| > T} \ket{\text{sign}_{\Delta I_x} = 0} \ket{a_1=0} \ket{\text{output}_x = 1} \\
&\quad + \ket{xy} \ket{I(x,y)} \ket{0}
\ket{|\Delta I_x| \leq T} \ket{\text{sign}_{\Delta I_x} = 0} \ket{a_1=0} \ket{\text{output}_x = 0} \Big) \\
&+ \sum_{\text{shifted}} \Big(
\ket{x}\ket{y} \ket{I(x,y)} \ket{0}
\ket{|\Delta I_x|} \ket{\text{sign}_{\Delta I_x} = 0} \ket{a_1=1}\ket{\text{output}_x=0} \\
&\quad + \ket{x+1}\ket{y} \ket{I(x+1,y)} \ket{0}
\ket{|\Delta I_x| > T} \ket{\text{sign}_{\Delta I_x} = 1} \ket{a_1=1}\ket{\text{output}_x=1} \\
&\quad + \ket{x+1}\ket{y} \ket{I(x+1,y)} \ket{0}
\ket{|\Delta I_x| \leq T} \ket{\text{sign}_{\Delta I_x} = 1} \ket{a_1=1}\ket{\text{output}_x=0} \Big)
\end{aligned}$ \\
\\
\hline
\\
6 &
$\begin{aligned}
\ket{\psi_6} &=  \Big(\sum_{\text{non-shifted}} \Big(
\ket{xy} \ket{I(x,y)} \ket{0}
\ket{|\Delta I_y| > T} \ket{\text{sign}_{\Delta I_y} = 0} \ket{a_2=0} \ket{\text{output}_y = 1} \\
&\quad + \ket{xy} \ket{I(x,y)} \ket{0}
\ket{|\Delta I_y| \leq T} \ket{\text{sign}_{\Delta I_y} = 0} \ket{a_2=0} \ket{\text{output}_y = 0} \Big) \\
&+ \sum_{\text{shifted}} \Big(
\ket{x}\ket{y} \ket{I(x,y)} \ket{0}
\ket{|\Delta I_y|} \ket{\text{sign}_{\Delta I_y} = 0} \ket{a_2=1}\ket{\text{output}_y=0} \\
&\quad + \ket{x}\ket{y+1} \ket{I(x,y+1)} \ket{0}
\ket{|\Delta I_y| > T} \ket{\text{sign}_{\Delta I_y} = 1} \ket{a_2=1}\ket{\text{output}_y=1} \\
&\quad + \ket{x}\ket{y+1} \ket{I(x,y+1)} \ket{0}
\ket{|\Delta I_y| \leq T} \ket{\text{sign}_{\Delta I_y} = 1} \ket{a_2=1}\ket{\text{output}_y=0} \Big)\Big)
\end{aligned}$ \\
\\
\hline
\\
7 &
$\ket{\text{output}} = \ket{\text{output}_x \vee \text{output}_y}$ \\
\\
\hline
\end{tabular}
\end{table*}

\subsection{Gradient Computation} \label{sec:gradient}

\begin{claim}[Directional Gradient Computation]
Let $\ket{x}\ket{y}\ket{0}^{\otimes q}\ket{0}^{\otimes q}$ denote the
position registers and two clean intensity registers.
The circuit described in this section implements the unitary mapping
\[
\ket{x}\ket{y}\ket{0}\ket{0}
\;\longmapsto\;
\ket{x}\ket{y}\ket{I(x,y)}\ket{I(x+1,y)},
\]
followed by a reversible transformation that produces the state
\[
\ket{\mathrm{sign}_{\Delta I_x}}\ket{\lvert I(x+1,y)-I(x,y)\rvert}.
\]
\end{claim}

The directional gradient computation submodule is split into two steps: neighborhood pixel generation and absolute value subtraction.

Given an image with dimensions $2^n \times 2^n$ with $q$ color values, we can construct a unitary oracle that takes input $\ket{x}\ket{y}\ket{0}^{\otimes{\log{q}}}$ and produces the state $\ket{x}\ket{y}\ket{I(x,y)}$. An efficient implementation of this oracle has been proposed by \cite{Amankwah_2022}, which has $O(q\times2^n)$ circuit depth. Applying an H$^{\otimes{n}}$ operator to both the $\ket{x}$ and $\ket{y}$ registers, then applying the NEQR oracle will yield the full image in NEQR. 

A ladder shift-up operator, $a^\dagger$, is used to modularly add $1$ to the $\ket{x}$ register, which has a depth of $O(n)$. The NEQR oracle is already configured to the original non-shifted image, so applying the oracle again, this time with a separate output register, the resulting state is:
\begin{align}
\ket{x+1}\ket{y}\ket{I(x,y)}\ket{I(x+1,y)}
\end{align}

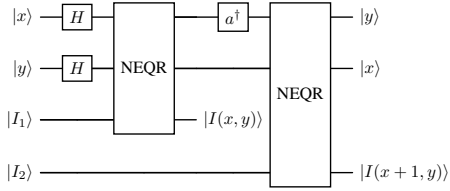
\begin{figure}[h]
    \centering
    \resizebox{0.7\columnwidth}{!}{
    \begin{quantikz}[column sep=0.45cm]
    \lstick{$\ket{x}$} & \gate{H} & \gate[3]{\text{NEQR}} &\qw& \gate{a^{\dagger}} & \gate[4]{\text{NEQR}}& \rstick{$\ket{y}$}\\
    \lstick{$\ket{y}$} & \gate{H} & \qw&\qw & \qw & \qw& \rstick{$\ket{x}$} \\
    \lstick{$\ket{I_1}$} & \qw &\qw   &\rstick{$\ket{I(x,y)}$}\\
    \lstick{$\ket{I_2}$} & \qw  & \qw  &\qw &\qw&\qw &\rstick{$\ket{I(x+1,y)}$} \\
    \end{quantikz}
    }
    \caption{Circuit Diagram for Generating Pixel Neighborhoods}
    \label{fig:shift_circuit}
\end{figure}

The circuit diagram for this stage is shown in Fig.~\ref{fig:shift_circuit}. Importantly, this step puts neighboring pixel intensity values in separate registers of the same quantum state, which is a necessary step for quantum arithmetic.

With both the original and shifted pixel intensities stored in separate quantum registers, the next step is to compute the magnitude of the gradient. Given two $q$-bit intensity values $I(x,y)$ and $I(x+1,y)$, the subtraction $\Delta{I_x}=I(x+1,y)-I(x,y)$ is computed using the quantum subtractor detailed in Subsec.~\ref{subsec:subtraction}.
 
It makes sense only to consider the magnitude of the gradient, so an absolute difference between neighboring pixels is required. Since we use signed integers, the most significant bit (MSB) of the difference $\Delta{I_x}$ is $1$ when the result is negative and $0$ if positive. To compute the absolute difference, another \texttt{S2C} is performed on all bits except the MSB of $\Delta{I_x}$, controlled when the MSB is $1$. The result is:
\begin{align}
\ket{\text{sign}_{\Delta{I_x}}}\ket{|I(x+1,y) - I(x,y)|}
\end{align}

The sign of the gradient represents the direction of the intensity shift, with a positive difference meaning the gradient goes from lower intensity to higher intensity, and a negative difference meaning the gradient goes from a higher to lower intensity. It is required to dedicate a qubit for the sign bit because it is crucial for the direction-aware edge shifting submodule.

\begin{figure}[h]
    \centering
    \resizebox{0.7\columnwidth}{!}{
    \begin{quantikz}[column sep=0.38cm]
    \lstick{$\ket{\alpha}$}  & \qw & \gate[3]{\text{Add}}& \qw& \rstick{$\ket{\alpha}$} \\
    \lstick{$\ket{\beta}$}  & \gate[2]{\text{S2C}} & \qw & \gate{\text{S2C}}&\rstick{$\ket{|\alpha-\beta|}$}  \\
    \lstick{$\ket{\text{Sign}_\beta}$} & \qw  & \qw   &\ctrl{-1}  &\rstick{$\ket{Sign}_{\alpha-\beta}$} \\
    \end{quantikz}
    }
    \caption{Circuit Diagram for Absolute-Value Subtraction}
    \label{fig:subtractor}
\end{figure}
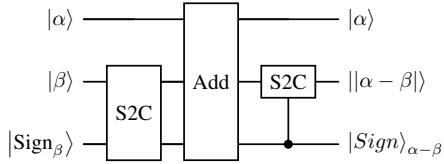

As an example, we compute $\ket{|\alpha-\beta|}$ for arbitrary $n$-qubit registers $\ket{\alpha}$, $\ket{\beta}$. The implementation of the absolute value subtractor for this problem is shown in Fig.~\ref{fig:subtractor}. 

The shifted position register is then shifted back down to its original position, and the adjoint of the NEQR oracle is performed on the shifted intensity register to reset it to the \ket{0} state so it can be reused for other directions.

\subsection{Direction-Aware Edge Shifting} \label{sec:edge_shift}

\begin{figure}[h]
    \centering
    \resizebox{0.9\columnwidth}{!}{
    \begin{quantikz}[column sep=0.5cm]
    \lstick{$\ket{I_1}$} & \qw & \qw & \gate[3]{\text{NEQR}^{\dagger}}& \qw& \gate[3]{\text{NEQR}} & \qw \\
    \lstick{$\ket{x}$} & \qw & \qw & \qw & \gate{a^{\dagger}}& \qw & \qw \\
    \lstick{$\ket{y}$}& \qw & \qw & \qw & \qw& \qw & \qw \\
    \lstick{$\ket{sign_{\Delta I_x}}$} & \ctrl{1}  & \gate{H}  &\ctrl{-3} &\ctrl{-2} &\ctrl{-3} & \qw \\
    \lstick{$\ket{a}$} & \targ{}  & \ctrl{-1}  & \qw&\qw &\qw & \qw
    \end{quantikz}
    }
    \caption{Direction-Aware Edge Shifting Circuit}
    \label{fig:direction-shifting}
\end{figure}
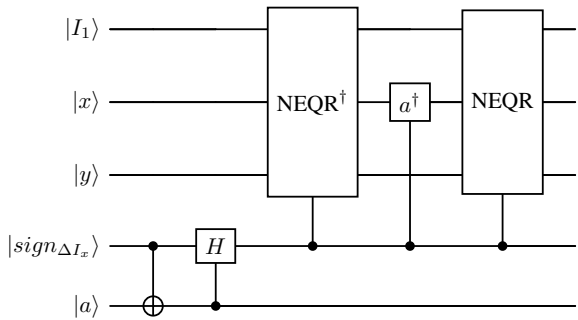

\begin{claim}[Direction-Aware Edge Shifting]
Given a pixel state
\[
\ket{x}\ket{y}\ket{|\Delta I_x|}\ket{\text{sign}_{\Delta I_x}},
\]
the circuit in Fig.~\ref{fig:direction-shifting} implements a reversible
mapping that:
\begin{enumerate}
\item Shifts the pixel position to $\ket{x+1}\ket{y}$ if
$\text{sign}_{\Delta I_x}=1$,
\item Leaves the position unchanged if $\text{sign}_{\Delta I_x}=0$,
\item Preserves a one-to-one correspondence between pixel locations and
basis states by duplicating shifted pixels and marking the unshifted
copy as a non-edge.
\end{enumerate}
\end{claim}

An edge typically lies at the boundary between a brighter pixel and a darker pixel, with the boundary being located at the darker pixel. The gradient computation module described in Section~\ref{sec:gradient} assigns the gradient magnitude $|\Delta I_x|$ to the pixel at $\ket{x}\ket{y}$, rather than the neighbor pixel at $\ket{x+1}\ket{y}$. As a result, when a light pixel precedes a dark pixel, the edge magnitude ends up being stored on the lighter pixel, effectively shifting the true edge position by one in the negative $x$ direction. To correct this and ensure edges are recorded at the darker (physically correct) location, we apply a conditional position shift based on the sign of the gradient. Specifically, if $\text{sign}_{\Delta I_x} = 1$ (meaning $I(x,y) > I(x+1,y)$), we shift the $x$-coordinate by one in the positive direction, thereby moving the marked edge into the darker pixel. If the sign is zero, no shift is applied. Formally, for each pixel, this module would like to perform the transformation: 

\begin{align}
\ket{x}\ket{y}
\ \rightarrow\ 
\begin{cases}
\ket{x+1}\ket{y}, & \text{if } \ket{\text{sign}_{\Delta I_x} = 1}, \\[4pt]
\ket{x}\ket{y}, & \text{if } \ket{\text{sign}_{\Delta I_x} = 0}.
\end{cases}
\end{align}

However, for the case when \ket{\text{sign}_{\Delta I_x} = 1}, \ket{x}\ket{y} transforms to \ket{x+1}\ket{y} while \ket{x-1}\ket{y} doesn't necessarily transform to \ket{x}\ket{y}, leaving us with a ``hole" in the $(x,y)$ pixel position. The next consideration is that there needs to be a quantum state associated with each pixel location in the image. To solve this problem, we put two ``copies" of the pixel information - one on the original location and the second on the shifted location. To ensure only the shifted pixel will become an edge pixel, we mark the pixel at the original location as not an edge. 

The implementation of this process is shown in Fig.~\ref{fig:direction-shifting}. The sign bit of each pixel selected for shifting is first placed in an equal superposition of \ket{0} and \ket{1} using a Hadamard gate. Since we would only like to Hadamard the sign qubit if it is in the \ket{1} state, we introduce an ancillary qubit. The computational basis state of the sign qubit is copied to the ancillary qubit, and it is used as a control for a Hadamard on the sign qubit. Then we use the sign qubit as a control qubit for a controlled ladder-up operator on the position register \ket{x}. 

For completeness, Fig.~\ref{fig:direction-shifting} also shows the surrounding controlled NEQR and $\text{NEQR}^\dagger$ blocks. These are not part of the shifting mechanism itself, but are used to re-encode shifted pixel intensities for when the circuit is applied across multiple gradient directions.

Pixels that are to be shifted begin in the state:
\begin{align}
    \ket{x}\ket{y}\ket{\text{sign}_{\Delta I_x}=1}\ket{a=0}
\end{align}
The sign qubit's computational basis state is copied to the ancillary:
\begin{align}
    \ket{x}\ket{y}\ket{\text{sign}_{\Delta I_x}=1}\ket{a=1}
\end{align}
The ancillary controls a Hadamard on the sign qubit:
\begin{align}
\frac{1}{\sqrt{2}}\Big(
\ket{x}\ket{y}\ket{\text{sign}_{\Delta I_x}=0}\ket{a=1} \\
- \ket{x+1}\ket{y}\ket{\text{sign}_{\Delta I_x}=1}\ket{a=1}\nonumber
\Big)
\end{align}

It is important to note that pixels not marked for shifting will have \ket{a=0}. The sign and ancillary qubits entangle the rest of the pixel information and provide a deterministic classification between the three possible cases this module considers, shown in Table~\ref{tab:shift_cases}.

\begin{table}[H]
\centering
\caption{Classification of pixel shifting cases}
\label{tab:shift_cases}
\begin{tabular}{c l}
\hline
\textbf{State} & \textbf{Case} \\
\hline 
$\ket{\text{sign}_{\Delta I_x}=0, a=0}$ & No shift\\ 
$\ket{\text{sign}_{\Delta I_x}=1, a=0}$ & Copied pixel shifted\\ 
$\ket{\text{sign}_{\Delta I_x}=1, a=1}$ & Copied pixel remains in-place\\ 
\hline
\end{tabular}
\end{table}

All pixels that are selected for shifting have a sign qubit in the \ket{1} state and those that do not have a sign qubit in the \ket{0} state. For the pixels that get selected for shifting, the copy that stays in its original location has \ket{a=0} and the copy that gets shifted has \ket{a=1}. Since the copied edges that remain in place are not to be considered as edges, we use their unique $\ket{\text{sign}_{\Delta I_x}=1, a=0}$ state as a non-edge mark in the thresholding step.

\subsection{Quantum Thresholding}

\begin{claim}[Quantum Thresholding]
Given a gradient magnitude register $\ket{|\Delta I|}$ and threshold $T$,
each thresholding method presented in this section implements a unitary
that marks pixels satisfying $|\Delta I| > T$. Specifically, this module performs the transformation: $\ket{0}\ket{|\Delta I|} \rightarrow \ket{|\Delta I| > T}\ket{|\Delta I|}$
\end{claim}

Once all gradient magnitudes along a particular direction are calculated and all pixels are shifted to the correct location, the next step is to classify which pixels represent actual edges. 

As described in Section~\ref{sec:edge_shift}, only two categories of pixels remain valid edge candidates: (a) pixels whose gradient sign is zero (no shift), and (b) those that have been shifted forward and marked with $\ket{a=1}$. The duplicated copy left at the original location when a pixel is shifted ($\ket{\text{sign}_{\Delta I}=1, a=0}$) is excluded from further consideration and treated as a guaranteed non-edge.

The other, and more significant, consideration is how large the magnitude of the gradient must be for a pixel to be considered an edge. This is achieved by comparing the gradient magnitude, \ket{|\Delta{I}|}, to a predefined threshold $T$. For any pixel that is considered to be an edge, if $|\Delta{I}| > T$, then it will be marked as an edge. 

Classically, this process would require sequentially iterating over each pixel and performing a conditional comparison. Comparing two $q$-bit integers takes $O(q)$ time in the classical RAM model, because you potentially check each bit in the worst case. Thus, for a $2^n \times 2^n$ image with $2^q$ color values, a classical thresholding algorithm would have a time complexity of $O(2^{2n}\cdot q)$. The need to perform such a comparison for every pixel introduces substantial overhead, which motivates the use of quantum parallelism to reduce the cost of this thresholding phase exponentially. This section will present a na\"ive arithmetic comparator and a novel approach that has $O(1)$ best-case performance with only one ancillary qubit.

\subsubsection{Comparator using the Quantum Ripple Carry Adder}

This method uses the same two's-complement subtraction logic as discussed earlier, but implements it with the Quantum Ripple-Carry Adder (QRCA) \cite{cuccaro2004newquantumripplecarryaddition}. In this approach, a clean register initialized to $\ket{T}$ is required, and the comparator is realized by subtracting $T$ from each gradient value in superposition:

\begin{align}
\ket{|\Delta I|}\ket{T} \;\mapsto\; \ket{|\Delta I|}\ket{T-|\Delta I|}.
\end{align}

The most significant bit of the difference serves as the sign bit, which determines if $|\Delta I| > T$.

However, the threshold register $\ket{T}$ must be explicitly stored to maintain reversibility because in the QRCA construction, both input registers are modified during computation. The MAJ gates overwrite intermediate values on the input wires, while propagating carry information forward. The UMA gates later use these modified values to uncompute the carries and restore the input values. Thus, the bits of $T$ are not simply classical controls; they are used as working qubits and are required throughout the full forward and reverse computation.

Thus, the QRCA-based comparator achieves linear depth $O(q)$ for a $q$-bit threshold, but requires an additional $q$-qubit register to store $\ket{T}$ in order to implement the subtraction reversibly.

\subsubsection{Quantum Partitioning Algorithm} \label{sec:qpa}

We now introduce the Quantum Partitioning Algorithm (QPA), which leverages phase kickback and a specialized Grover Phase Oracle \cite{Grover1996}, which we call the Fast Threshold Phase Oracle (FTPO). This method avoids the need for an additional threshold register and significantly reduces the circuit depth compared to arithmetic-based approaches for most practical cases.

The key idea is to encode the comparison outcome directly into the phase of an ancillary qubit, rather than explicitly computing the difference $T - |\Delta I|$. The FTPO flips the phase of every basis state $\ket{s}$ in the ensemble of the intensity register $\ket{|\Delta I|}$ that is greater than the value of $T$ while leaving all other states unchanged:

\begin{align}
\ket{s} \xrightarrow{\;\text{FTPO}\;} 
\begin{cases}
\ket{s}, & s < T, \\
-\ket{s}, & s \geq T.
\end{cases}
\end{align}

The na\"ive Grover approach to mark all $\ket{s}: s>T$ is to brute force every such $s$ using multi-controlled Z gates \cite{javadiabhari2024quantumcomputingqiskit}. For a $q$-bit threshold, this approach requires $O(2^q)$ multi-controlled Z gates, each with $O(q)$ control qubits. Each of these multi-controlled Z gates can be constructed in $O(\log{q})$ depth in the Clifford + T model, using one ancillary qubit \cite{nie2024quantumcircuitmultiqubittoffoli}. Thus, this phase oracle implementation has a depth of $O(2^q\log q)$.

Our novel FTPO avoids this exponential overhead by exploiting the hierarchical bit ordering of binary numbers. Instead of individually comparing each bit, the FTPO recursively partitions the Hilbert space into groups of states that are guaranteed to have $s > T$ based on prefix matching. To see this, let us consider the construction formally.

We want to mark all $q$-bit integers $s$ such that $s > T$, where $T = (t_{q-1}t_{q-2}\ldots t_0)$.\\ \\
\textbf{Base Case.}

Suppose that the most significant bit of $T$ is $0$. Any integer $s$ with MSB equal to $1$ must satisfy $s > T$, and we mark such states $\ket{s}$ by flipping its phase.\\ \\
\textbf{Inductive Step.}

We then remove the most significant bit and consider the remaining suffix $T'$. If the MSB of $T'$ is $1$, no additional states are guaranteed to exceed $T$, and we proceed recursively. If instead the MSB of $T'$ is $0$, then any integer whose prefix matches that of $T$ but flips this bit to $1$ must exceed $T$. We flip the phase of all such $\ket{s}$.\\

At each step, the prefix before $T'$ is fixed, ensuring that the subsets of states marked at different recursion levels are disjoint. This uniqueness property is important because if the phase of a state is flipped an even number of times, then it will return to a positive phase, effectively not being marked at all. Thus, by induction, every state $s > T$ is marked exactly once.

The recursive construction of the FTPO can be implemented in a quantum circuit using a series of multi-controlled Z gates. We iterate through the bits of $T$, starting with the MSB. For each bit position $i$ in the threshold $T$ where $t_i = 0$, a multi-controlled Z gate is appended with controls on all qubits more significant than $t_i$. The presence of the control bits is to ensure that the prefixes of the input state and the threshold match. This is done by applying an X gate on all prefix qubits after its recursive level is executed, and then uncomputing the X gate at the end of the FTPO.

\begin{algorithm}[h]
\caption{Fast Threshold Phase Oracle (FTPO) circuit construction}
\KwIn{Threshold bitstring $T = (t_{q-1}t_{q-2}\ldots t_0)$}
uncompute\_list $\gets$ [ ]\;
\For{$i \gets q-1$ \KwTo $0$}{
    \If{$t_i = 0$}{
        Apply multi-controlled $Z$ on qubit $i$, controlled by qubits $j>i$\;
        Apply $X$ gate to $t_i$\;
        Append $t_i$ to uncompute\_list\;
    }
}
Apply $X$ gates to all qubits in uncompute\_list\;
\end{algorithm}

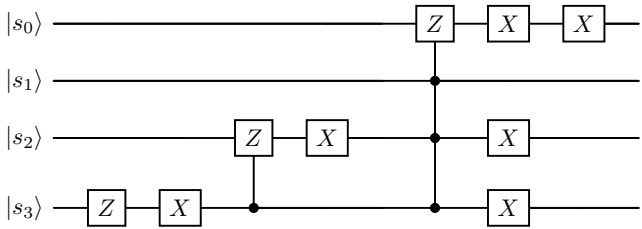
\begin{figure}[h]
\centering
\resizebox{\columnwidth}{!}{
\begin{quantikz}[column sep=0.5cm] 
\lstick{$\ket{s_0}$} & \qw & \qw & \qw &\qw & \qw & \gate{Z} & \gate{X} & \gate{X} &\qw\\
\lstick{$\ket{s_1}$} & \qw & \qw & \qw &\qw & \qw & \ctrl{-1} & \qw & \qw &\qw\\
\lstick{$\ket{s_2}$} & \qw & \qw & \gate{Z} &\gate{X} & \qw & \ctrl{-2} & \gate{X} & \qw&\qw  \\
\lstick{$\ket{s_3}$} & \gate{Z} & \gate{X} & \ctrl{-1} & \qw & \qw & \ctrl{-3} & \gate{X} & \qw&\qw 
\end{quantikz}
}
\caption{FTPO Implementation for $T=0010$}
\label{fig:ftpo_example}
\end{figure}

For a visual example, the implementation of the FTPO for threshold $T=0010$ is shown in Fig.~\ref{fig:ftpo_example}

Once constructed, the FTPO is invoked only once in the QPA to perform thresholding over the gradient magnitudes. We introduce one ancillary qubit, initialized to the $\ket{+}$ state by using a Hadamard gate. The procedure begins with the state:
\begin{align}
\ket{\psi_1} = \frac{1}{\sqrt{2}}(\ket{0} + \ket{1}) \otimes \ket{|\Delta I|},
\end{align}
where \ket{|\Delta I|} encodes the ensemble of all intensity gradient magnitudes. The FTPO acts as a Grover-style phase oracle, flipping the phase of all states $\ket{s}$ with $s > T$, conditioned on the control qubit:
\begin{align}
\ket{\psi_2} = \frac{1}{\sqrt{2}}(\ket{0} + \ket{1})\ket{W^\perp} + \frac{1}{\sqrt{2}}(\ket{0} - \ket{1})\ket{W},
\end{align}
where 
\begin{align}\label{eq:W}
    \ket{W} = \left\{ \ket{\psi} : \psi \in \mathbb{N}, \psi > T \right\}
\end{align}
and
\begin{align}\label{eq:Wperp}
    \ket{W^\perp} = \left\{ \ket{\psi} : \psi \in \mathbb{N}, \psi \leq T \right\}
\end{align}
Applying a Hadamard to the control qubit uncomputes the superposition in the ancillary qubit:  
\begin{align}
\ket{\psi_3} = \ket{0} \ket{W^\perp} + \ket{1} \ket{W}.
\end{align}
which achieves the desired partition of the state space into pixels above and below the threshold.

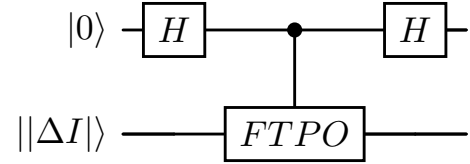
\begin{figure}[h]
\centering
\resizebox{0.75\columnwidth}{!}{
\begin{quantikz}[column sep=0.2cm] 
\lstick{$\ket{0}$} & \gate{H} & \ctrl{1} & \gate{H} & \qw \\
\lstick{$\ket{|\Delta I|}$} & \qw & \gate{FTPO} & \qw & \qw
\end{quantikz}
}
\caption{QPA Circuit Implementation using FTPO}
\label{fig:ftpo_threshold}
\end{figure}

The implementation of the Quantum Partitioning Algorithm using the FTPO is shown in Fig.~\ref{fig:ftpo_threshold}.

Our Quantum Partitioning Algorithm only requires one ancillary qubit, and one function call to the FTPO. The FTPO applies a multi-controlled Z gate for every $0$ in the threshold $T$. The number of control qubits in each of these gates is dependent on the bit position $t_i$ of the $0$. 

For a $q$-bit threshold, the worst case is if the bits of $T$ are all $0$'s, leading to $q$ multi-controlled Z gates. Decomposing into Clifford + T, this yields a worst-case depth of
\begin{align}
    \sum_{i=0}^{q-1}O(\log{i}) \in O(q\log{q)}.
\end{align}

However, in most practical settings, the threshold corresponds to a fraction of the maximum range, typically a power-of-two fraction such as one-half, one-quarter, or one-eighth. As shown in \eqref{eq:W} and \eqref{eq:Wperp}, a threshold bitstring $T$ with $n$ leading zeros followed by ones partitions the set in a way that the upper subset $W$ contains exactly a $1 - 1/{2^n}$ fraction of the total states, while the lower subset $W^\perp$ contains the remaining $1/{2^n}$. In practice, thresholds for image processing won't need more than a constant number of leading zeros, so only a constant number of multi-controlled Z gates are required, each with a constant number of control qubits. Thus, for most applications, the time complexity of the QPA is effectively $O(1)$.

For edge detection, this assumption of approximate thresholds is both reasonable and practical. Hence, our thresholding algorithm achieves constant-time performance in this setting, while maintaining an auxiliary space cost of only one ancillary qubit. This makes it the ideal choice for the quantum thresholding module presented in this work. 

\section{Results and Discussion} \label{sec:results}

This section will evaluate the performance of our proposed algorithm against existing classical and quantum edge detection algorithms. 

\subsection{Algorithm Complexity Analysis}

Consider the input to be a $2^n \times 2^n$ image with $2^q$ color intensity values per pixel. \\

\subsubsection{Time Complexity}

The complexity of image preparation in NEQR form is $O(q\times2^n)$ \cite{Amankwah_2022}. Although this step has a high overhead, NEQR preparation is not considered part of quantum image processing and is typically ignored from runtime analyses \cite{Liu_2022}\cite{chetia2021}\cite{fan2019}. Copying the computational basis state of an intensity register is implemented with one CNOT gate on each qubit position, which has a complexity of $O(q)$. The ladder shift operators are implemented using the QRCA, which has a complexity of $O(n)$. Similarly, the absolute value subtractor is decomposed into three QRCAs and a constant number of elementary gates, yielding a complexity of $O(n)$. The direction-aware edge shifting module is composed of elementary gates, a controlled ladder operator, and two controlled image encoding oracles. Ignoring the effect of the NEQR oracles, this module has a complexity of $O(n)$. Finally, the quantum thresholding module operates in constant time, under the practically relevant assumption of approximate thresholds, as enabled by the Quantum Partitioning Algorithm. In the exceptional cases where this assumption does not hold, the thresholding step incurs a cost of $O(q\log{q})$. Thus, our algorithm has an overall time complexity of $O(n+q)$.

Although it is not considered in the complexity analysis, the intermediate NEQR oracle overhead can be eliminated by using additional computation registers to avoid the need to reset the intensity registers for reuse.\\

\subsubsection{Space Complexity}

As shown in Fig.~\ref{fig:algo_circuit}, the registers that are measured (i.e., outputs of the algorithm) are \ket{x}, \ket{y}, and \ket{output}. The additional computational registers: \ket{I_1}, \ket{I_2}, and \ket{GradX} are each $q$-qubits. The other qubits used, whether for storage or ancillaries, are all constant-sized. Thus, the auxiliary space used by this algorithm is $O(q)$. Furthermore, the output edge map of this algorithm is $O(n)$, which exponentially compresses a classical image, requiring $O(2^q\cdot2^{2n})$ bits to store.

\subsubsection{Comparison with Existing Classical and Quantum Algorithms}

Table~\ref{tab:complexity} compares the time complexity of our proposed algorithm with both classical edge detection methods (Sobel, Canny) and existing quantum approaches. While QHED has a faster runtime compared to our algorithm, this does not include the heavy classical postprocessing steps required to produce an accurate edge map output. Table~\ref{tab:space_complexity} shows that our method requires fewer ancillary qubits, reducing the count by a linear factor in the number of intensity qubits relative to the leading quantum edge detection algorithm. 

\begin{table}[h]
    \centering
    \caption{Time complexity comparison of edge detection algorithms.}
    \label{tab:complexity}
    \begin{tabular}{lc}
        \hline
        \textbf{Algorithm} & \textbf{Time Complexity}\\
        \hline
        Sobel\cite{Sobel} & $O(2^{2n})$ \\
        Canny\cite{Canny} & $O(2^{2n})$ \\
        QHED \cite{yaoqhed} & $O(n^2)$\\
        NEQR QSobel\cite{Liu_2022} & $O(n^2+q^2)$ \\
        Our Algorithm & $O(n+q)$ \\
        \hline
    \end{tabular}
\end{table}

\begin{table}[h]
    \centering
    \caption{Computational qubit count of quantum edge detection algorithms.}
    \label{tab:space_complexity}
    \begin{tabular}{lc}
        \hline
        \textbf{Algorithm} & \textbf{Number of Temporary Qubits}\\
        \hline
        NEQR QSobel\cite{Liu_2022} & $9q + O(1)$ \\
        Our Algorithm & $3q + O(1)$ \\
        \hline
    \end{tabular}
\end{table}

\subsection{Experimental Analysis}
\begin{figure}[h]
    \centering
    \includegraphics[width=1\linewidth]{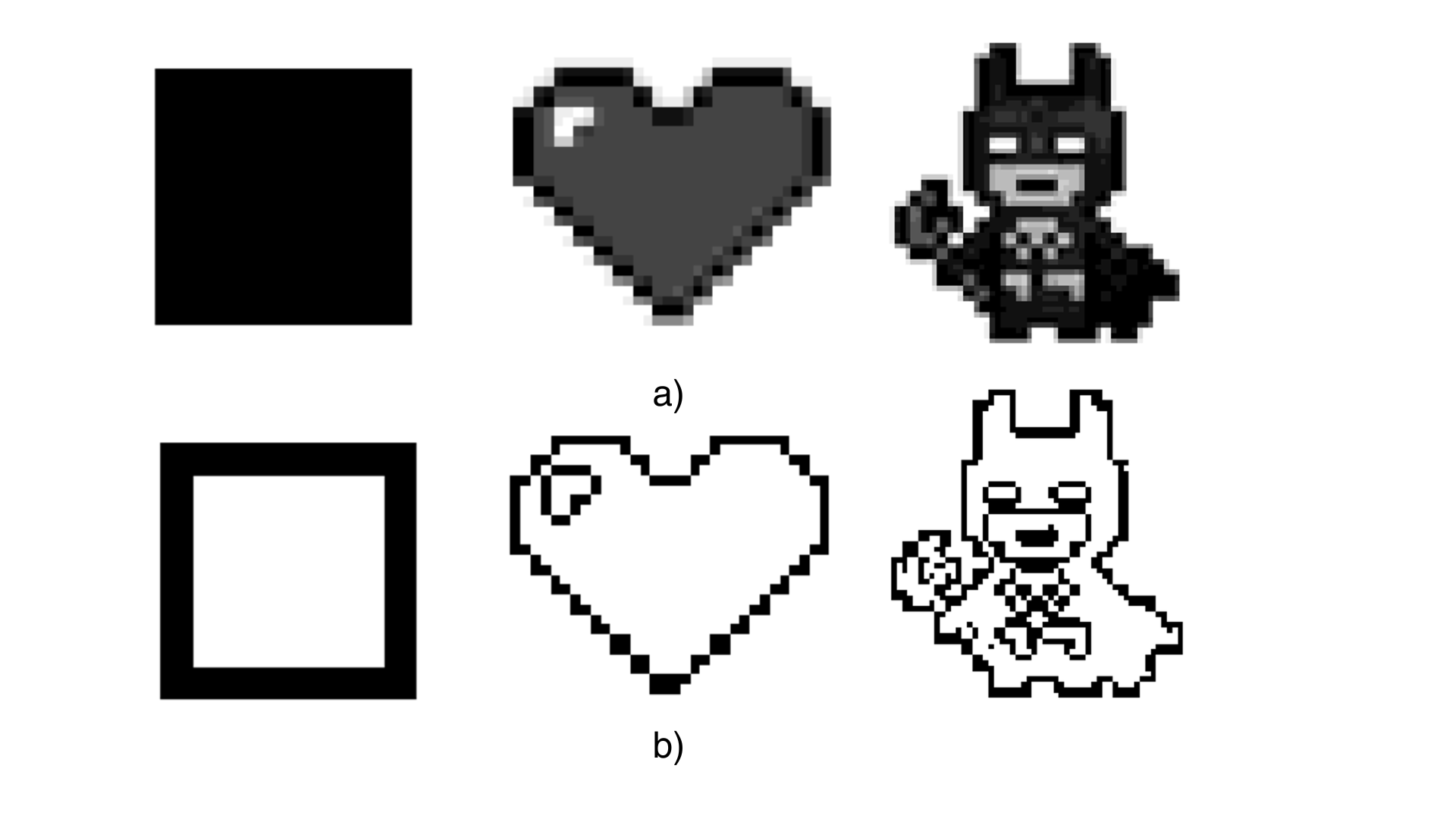}
    \caption{a) Three grayscale test images. b) The resulting edge maps our algorithm outputs.}
    \label{fig:results}
\end{figure}

To evaluate the effectiveness of our proposed algorithm, we implement and test our algorithm using the IBM Qiskit framework on a local simulator \cite{javadiabhari2024quantumcomputingqiskit}. Figure~\ref{fig:results} displays three grayscale images along with the corresponding edge maps that this algorithm outputs. As shown by Figure~\ref {fig:results}, our algorithm accurately captures edge boundaries at the pixel-level while ignoring noisy edges.

For larger images, the uncomputation steps required to reset intermediate registers are performed classically to reduce simulation overhead while preserving logical correctness. These steps have been explicitly verified on small images to ensure the resetting operations behave as intended. To construct the output edge map, we measure the designated output qubit in the $\ket{1}$ state and record the corresponding position qubits, which collectively encode the detected edge locations.

\section{Conclusion} \label{sec:conclusion}

In this work, we have presented a fully quantum gradient-based edge detection algorithm that leverages NEQR encoding, quantum arithmetic circuits, cyclic image shifting, a direction-aware edge shifting procedure, and a novel quantum thresholding algorithm. Beyond image processing, the Quantum Partitioning Algorithm itself has broader applicability: its ability to partition an unordered set in best-case $O(1)$ complexity, with only constant auxiliary space, offers significant potential for problems requiring efficient filtering of quantum states. 

Our analysis demonstrates that the algorithm achieves a time complexity of $O(n+q)$ in practical settings while only requiring $3q + O(1)$ computational qubits for a $2^n\times 2^n$ image with $2^q$ intensity values per pixel, a significant reduction compared to prior quantum edge detection approaches \cite{yaoqhed}\cite{Shubha2024pex}\cite{Liu_2022}. Moreover, our algorithm achieves a quantum advantage compared to leading classical approaches, exponentially compressing the output edge map while also exponentially speeding up the edge detection process \cite{Sobel}\cite{Canny}.

Experimental validation on test images confirms our algorithm's ability to reliably identify meaningful intensity transitions and produce edge maps with low noise, consistent with our expectations. These results demonstrate the feasibility of performing complex image processing tasks entirely within the quantum regime, without intermediate measurements or classical processing. 

Future work would optimize the NEQR image encoding steps for faster state preparation and register resetting. Furthermore, multi-control gate optimization would greatly improve the worst-case runtime of our QPA, which is heavily reliant on multi-control Z gates \cite{nie2024quantumcircuitmultiqubittoffoli}. Finally, there are many possibilities for connecting quantum machine learning (QML) with our algorithm. For example, a possible application of QML is to train a model to determine an optimal threshold $T$ for certain classes of images (e.g., medical image scans, or object detection for autonomous vehicles). Since our algorithm outputs an edge map as a quantum state, it can enable powerful hybrid or fully-quantum workflows for computer vision tasks. 

Overall, this research provides a more practical and efficient approach to quantum edge detection compared to previous works, while highlighting the potential of quantum image processing in the NISQ era.

\section*{Acknowledgment}

The author gives many thanks to Luke Schaeffer for the many helpful discussions and reviews of this manuscript. This work was funded in part by Transformative Quantum Technologies.

%\begin{acknowledgments}
%This work was funded in part by Transformative Quantum Technologies and by the President's %Research Award, University of Waterloo.
%\end{acknowledgments}

% Produces the bibliography via BibTeX.

\end{document}